\documentstyle[12pt]{article}

\begin{document}

\begin{center}
{\bf Searching for the Quark-Gluon Plasma}
\vspace{2cm}

Sherman Frankel

Physics Dept, University of Pennsylvania

Philadelphia, PA 19104

October 1, 1994

\vspace{3cm}

Abstract

\end{center}

The claims for production of high energy densities and possible
new states of
matter in collisions of nuclei by George F. Bertsch (Science, {\bf 265}
(1994) 480-481) are examined and compared with simple
   explanations of
the data which have appeared in the literature. We point out that:
Energy densities in
S-S collisions are almost identical with those in p-p collisions; J/Psi
Feynman x distributions are reproduced by conventional application of
energy loss considerations; The effective sizes of collisions obtained
from Bose-Einstein correlation measurements appear to be ``large''
because of the accepted view that quark-antiquark pairs take time to
dress themselves into on-shell bosons.

\newpage
         George Bertsch's ``perspective'' in {\it Science} with this title
               (July 22, 1994)
         suggests that there are new scientific
measurements that hold out the possibility that one might actually see
evidence of the production of a ``quark-gluon plasma'' in nucleus-nucleus
collisions   at high energies. In such a plasma the quarks and gluons
that are normally confined within the protons and neutrons are free to
escape and wander throughout the nuclear volume.

        While one might like to believe that the measurements he
describes are suggestive of new phenomena, simple considerations such as
conservation of energy and momentum appear to fit the data without any
need for ``new physics''.

                     Bertsch's popular account
suggests that nucleus-nucleus collisions are reaching especially
high energy
densities. However the data he refers to
do not at all give evidence for
this conclusion. For example, the Sulphur-Sulphur and proton-proton
            reactions to which he refers \cite{Alber} show
particle densities that hardly differ. The very slight differences in
energy density distributions
are understood from simple kinematics.
 Also, one can understand the remark that
``many of the nucleons are slowed almost essentially to a stop''
only if one describes
``stopping''as moving with a speed  somewhat less than
the speed of light!

Bertsch
   refers to the J/Psi suppression in nuclear collisions as further
evidence for a hot dense environment. Yet workers have shown
years ago \cite{Gavin}\cite{Frankel1}
 \cite {Gerschel} \cite{Frankel2}
that
there is nothing mysterious about the so-called "suppression",
which is the reduction of yields of J/Psi's in nuclear interactions as
compared with  proton-proton interactions.
Since the incoming hadrons usually
slow down to lower energies as the result of ``soft'' collisions
before they
produce a J/Psi, and since the probability of producing a J/Psi is known
to decrease markedly
with bombarding energy,
                         and since
outgoing J/Psi's even slow down and disappear
          when they bump
into ``spectator'' nucleons in the nucleus,
the ``suppression'' has conventional explanations.
This is also shown dramatically by the fact that the
J/Psi's in nuclear collisions are observed to have  very
different velocity distributions
    as a function of nuclear size  \cite{Katsenevas}
which is
due to simple energy loss effects \cite{Frankel2}.

      Prof. Bertsch  mentions  the results of beautiful experiments on
the interference of meson waves in nucleus-nucleus collisions which
he claims show that the produced pions may
exist as another form of matter for a time
about
twice as long as the collision lasts and before that form
turns into real mesons. There are conventinal reasons to expect the
``sizes''
to be increased. For example,
       scattering of pions in the hadronic matter takes
time and the measurements measure pion properties after the last scatter
and the pions have moved away from the basic interaction region.
          Pions consist of pairs of quarks and antiquarks surrounded
by several gluons.
It is a basic tenet
that ``bare''quark-antiquark pairs are produced directly in
high energy  proton and neutron collisions.
Accepted  theory requires that it take some time for the
quarks to become ``dressed'', i.e., to become
surrounded by gluons  and evolve into the  real pions that the experiment
detects.
During that formation
time the pions can move outward to larger radii
so the measured size of the interaction
is naturally larger.
Thus a somewhat larger size is even expected in p-p collisions.
Actually the interpretation of these interference experiments in terms of
a ``size'' is itself a fuzzy concept.

It was not at the end but at the beginning
of the 1980's                     that the first
searches for the quark-gluon plasma in nucleus-nucleus collisions were
made. These were at energies more than twice that of present accelerators
and took place at the now-dismantled
    ISR accelerator at CERN. Although the nuclei were light (Helium),
the energy densities reached in the observed
very high multiplicity events, calculated
from either the Bjorken or Gyulassy and Matsui estimates,
were larger than the needed threshhold.
   No
evidence for the QGP  appeared,
even in precise multiplicity
measurements \cite{Callen}, or strange particle production
\cite{Frati}.

The quark-gluon plasma is believed to be formed
when nucleons are pressed together at extraordinarily high
pressures, such as existed in the early universe.
``Lattice gauge'' calculations, which simulate this static high pressure,
indicate that the appropriate phase transition could indeed take place.
Unfortunately, as Bertsch remarks, the collisions between nuclei take
place ``for only a short time''
so the transition may not have time to take place. (The collision
time is about $10^{-23}$ seconds.)
                                   After expenditures of large sums
of money over several decades
there are still no signs of the quark-gluon plasma.
The lesson being given  may be  that it is not
likely
that one can make pressed-nuclearduck by throwing two nuclearducks
at each other.

\flushright Bertsch  print date: \today  UPR 637T
\end{document}